\documentclass[aps,groupedaddress,showpacs]{revtex4}
\usepackage{amsmath}
\usepackage{amssymb}
\usepackage{amsfonts}
\usepackage[dvips]{graphicx}
\usepackage[]{caption}
\usepackage[]{epsfig}
\begin{document}

\newcommand{\col}{colloid }
\newcommand{\pol}{polymer }
\newcommand{\inter}{interacting }
\newcommand{\dep}{depletion }
\newcommand{\rg}{R_{g}}
\newcommand{\hs}{hard sphere }

\title{Thermodynamic perturbation theory of the phase behaviour of colloid / interacting polymer mixtures}

\author{B. Rotenberg}
\affiliation{Department of Chemistry,
Lensfield Road,
Cambridge CB2 1EW,
United Kingdom}
\author{J. Dzubiella}
\email[e-mail address: ] {jd319@cam.ac.uk}
\affiliation{Department of Chemistry,
Lensfield Road,
Cambridge CB2 1EW,
United Kingdom}
\author{J.-P. Hansen}
\affiliation{Department of Chemistry,
Lensfield Road,
Cambridge CB2 1EW,
United Kingdom}
\author{A. A. Louis}
\affiliation{Department of Chemistry,
Lensfield Road,
Cambridge CB2 1EW,
United Kingdom}
\date{\today}

\begin{abstract}
We use thermodynamic perturbation theory to calculate the free
energies and resulting phase diagrams of binary systems of spherical
colloidal particles and interacting polymer coils in good solvent
within an effective one-component representation of such mixtures,
whereby the colloidal particles interact via a polymer-induced
depletion potential. MC simulations are used to test the convergence
of the high temperature expansion of the free energy. The phase
diagrams calculated for several polymer to colloid size ratios differ
considerably from the results of similar calculations for mixtures of
colloids and ideal (non-interacting) polymers, and are in good overall
agreement with the results of an explicit two-component
representation of the same system, which includes more than two-body
depletion forces.
\end{abstract}

\pacs{05.70.Ln, 61.20.Ja, 82.70.Dd, 64.70.Dv}

\maketitle

\section{Introduction}

The structure, rheology and phase behaviour of sterically stabilized
colloidal dispersions are strongly affected by the presence of non
adsorbing polymer. Nearly fifty years ago Asakura and Oosawa realized
that finite concentrations of polymer coils would induce an effective
attraction between colloidal particles, of essentially {\it entropic}
origin, the so-called depletion interaction \cite{AO}. Since the
initial colloid-polymer Hamiltonian involves only repulsive
interactions between all pairs of particles, the polymer-induced
effective attraction between colloids, which results from tracing out
the polymer degrees of freedom, was referred to as ``attraction
through repulsion'' by A. Vrij. For non-interacting (ideal) polymers,
the range of the depletion attraction is independent of polymer
concentration, and close to the polymer radius of gyration $R_{g}$,
while the depth of the attractive well, when two colloids touch, is
proportional to polymer concentration. Consequently, one expects that
for sufficiently high concentration, and for not too small size ratio
$q=R_{g}/R_{c}$ (where $R_{c}$ is the radius of the spherical
colloids), the effective attraction may drive a depletion-induced
phase separation into colloid-rich (``liquid'') and colloid-poor
(``gas'') colloidal dispersions, similar to condensation in simple
fluids. This phase transition, which is in fact a colloid-polymer
demixing transition, was investigated by Gast {\it et al.}
\cite{gast}, who first calculated the phase diagram from thermodynamic
perturbation theory. Their findings were later confirmed by the
Monte-Carlo (MC) simulations of Meijer and Frenkel \cite{meijer}, and
the free volume theory of Lekkerkerker {\it et al.} \cite{lekk}.  The
predicted phase diagrams agree qualitatively with experimental
findings for various colloid / polymer mixtures
\cite{experiments,ilett}.

More recently the question was raised of how interactions between
polymer coils would affect the phase behaviour compared to that of
ideal polymers \cite{warren,fuchs,PRL89,aarts,schmidt}. The early
theoretical investigations into the problem were made at the
two-component level, involving an explicit consideration of the
polymer coils. However very recently the depletion-driven effective
pair-potential between two colloidal particles in a bath of
interacting polymers in good solvent, modelled as self-avoiding walk
(SAW) polymers, was calculated by MC simulations \cite{JCP117}. A
simple analytic form, with coefficients determined by the SAW polymer
osmotic equation of state and surface tension, yields excellent
agreement with the simulation data over a wide range of polymer to
colloid size ratio $q$, and polymer concentrations \cite{JCP117}.

In this paper we use thermodynamic perturbation theory \cite{simplliq} to calculate the phase diagram of mixtures of
hard sphere colloids and interacting polymers within the effective one-component representation, whereby the colloidal
particles interact via the above-mentioned depletion potential induced by the SAW polymers. The results of these 
calculations can be directly compared to the predictions of recent MC 
simulations of the two-component representation of the same system \cite{PRL89}, which agree quantitatively with recent
experiments \cite{ramak}.
Any discrepancies between the phase diagrams obtained within the effective one-component and
two-component representations can then be traced back to more-than-two-body depletion interactions between colloidal
particles, which are automatically included in the latter representation, but are of course neglected in the pairwise
additive effective one-component picture considered in the present paper. 

Thermodynamic perturbation theory was previously applied to mixtures
of hard sphere colloids and ideal polymers within the effective
one-component representation by Gast {\it et al.} \cite{gast}, and
improved by Dijkstra {\it et al.}  \cite{dijkstra}. Similar
calculations were recently published for mixtures of colloids and star
polymers for several functionalities $f$ (where $f$ is the number of
identical arms of the star polymer connected at the centre), again
within an effective one-component representation, as well as within
the two-component description \cite{dzubiella}. Star polymers are
particularly instructive depletants, since upon varying the
functionality, they change continuously from linear polymer ($f=2$) to
hard-sphere like behaviour ($f \rightarrow \infty$). The phase
diagrams obtained in ref.  \cite{dzubiella} for $f=2$ are thus, in
principle directly comparable to the results presented in this
paper. Such a comparison will be made in section $4$, but is only
tentative, since the calculations in ref. \cite{dzubiella} neglect the
polymer concentration-dependence of the effective colloid-polymer and
the polymer-polymer interactions which are not negligible
\cite{PRL89,JCP114,macromol}. \\

\section{One and two-component representations}

\begin{figure}[here]
\begin{center}
\includegraphics[width=12cm,height=6cm]{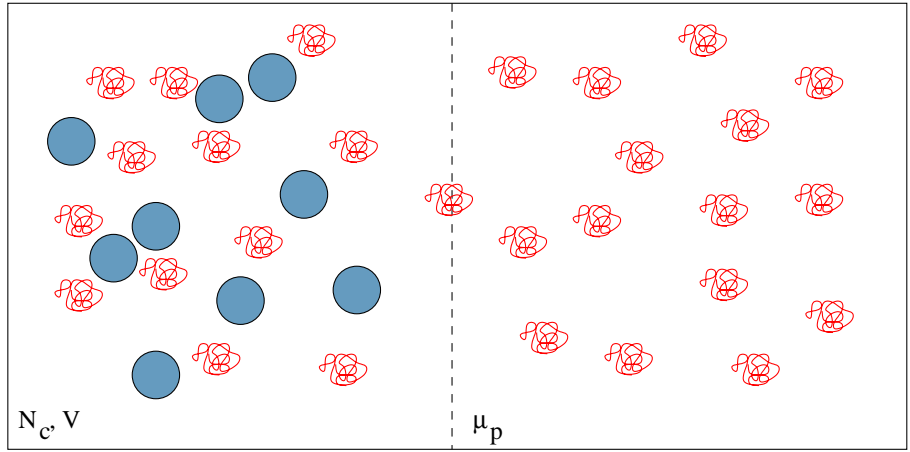}
\end{center}
\caption{Schematic representation of the semi-grand canonical ensemble. $N_{c}$ colloidal particles in a volume $V$ are in
osmotic equilibrium with a polymer reservoir of fixed chemical potential $\mu_{p}$.}
\label{semigrand.fig}
\end{figure}

Consider a system of $N_{c}$ spherical colloidal particles of radius $R_{c}$, in a bath of linear polymers which are in
equilibrium with a polymer reservoir of fixed chemical potential $\mu_{p}$. The corresponding semi-grand canonical
description is schematically represented in figure \ref{semigrand.fig}. The colloidal particles interact via the standard
hard sphere potential, while each polymer is made up of $L$ monomers or segments; segments from the same or different
chains are not allowed to overlap. In good solvent, this excluded volume constraint is the only monomer-monomer interaction
and for sufficiently large $L$, where chemical details become irrelevant, 
the interacting polymers may be accurately modelled by
self avoiding walks on a three-dimensional lattice. The monomers, moreover, are not allowed to penetrate the hard sphere
colloids. At finite colloid and polymer concentrations, such a binary mixture of hard spheres and interacting polymers
poses a formidable problem to theoreticians and simulators alike. 

One coarse-graining strategy which has proved very successful is to
trace out individual monomer degrees of freedom in the ``polymers as
soft colloids'' paradigm, whereby the total interaction between two
polymer coils, averaged over all monomer configurations, reduces to an
effective (entropic) interaction between their centres of mass, which
depends on polymer concentration \cite{JCP114}. Similarly, one can
trace out the monomer degrees of freedom, to derive a state-dependent
effective interaction between the hard sphere colloids and the centres
of mass of the polymer coils \cite{JCP114,macromol}. This
coarse-graining procedure, which amounts to a reduction of the number
of degrees of freedom of each polymer from $\mathcal{O}(L)$ to $3$,
leads to a two-component representation of ``hard'' and effective
``soft'' colloids, which has been exploited in recent MC simulations
to determine the phase diagram of colloid / interacting polymer
mixtures for several size ratios $q$ \cite{PRL89}. However, following
the Asakura-Oosawa (AO) strategy for non-interacting polymers, one can
carry the coarse-graining procedure one step further, by eliminating
the polymer degrees of freedom altogether and determining the
resulting depletion interactions between the colloidal particles. If
this procedure is carried out in the semi-grand canonical ensemble,
the total effective interaction energy between $N_{c}$ colloidal
particles for any configuration $\overrightarrow{r}^{N_{c}}$ is:
\begin{equation}\label{veff}
V_{cc}^{eff}(\vec{r}^{N_{c}}) \; \equiv \; V_{cc}(\vec{r}^{N_{c}}) 
- k_{B} \, T \; {\rm ln}<{\rm exp}[-\beta V_{cp}]>_{\mu_{p},V,T}(\vec{r}^{N_{c}})
\end{equation}
\noindent
where $V_{cc}$ is the direct colloid-colloid interaction energy, while $V_{cp}$ is the total colloid-polymer interaction;
the brackets denote a grand-canonical average over polymer degrees of freedom at fixed polymer chemical potential, volume
and temperature, and a given colloid configuration. In the model under consideration, $V_{cc}$ and $V_{cp}$ are the
colloid-colloid and colloid-monomer excluded volume interactions, which are pair-wise additive. The depletion interactions
between the $N_{c}$ colloidal particles are given by the second term on the r.h.s. of Eq. (\ref{veff}) and are not,
a priori, pair-wise additive. However, for sufficiently low colloid concentration, or small size ratio $q$, 
the pair-wise additive contribution
dominates. 
\subsection{Depletion potential for interacting polymers}
The depletion pair potential between an isolated pair of colloidal spheres has been determined as a function of
the centre-to-centre distance, and over a range of interacting polymer concentrations covering the dilute and semi-dilute 
regimes, in ref. \cite{JCP117}. The resulting depletion pair potential is accurately reproduced by the following simple
semi-empirical form, inspired by the Derjaguin approximation \cite{JCP117}:
\begin{equation}
\label{louis.eq}
\beta v_{s}(r) = 
\begin{cases}
- \pi \, R_{c} \, \gamma_{w}(\rho) \, D_{s}(\rho)  \left( 1- \frac{x}{D_{s}(\rho)} \right)^{2}
& ;  \quad 0 \; \le x \; \le D_{s}(\rho) \; , \\
0 \quad {\rm otherwise}
\end{cases}
\end{equation}
\noindent
$r$ is the centre-to-centre separation, while $x = r - 2R_{c}$ is the surface-to-surface distance between the colloids;
$\gamma_{w}(\rho)$ is the polymer surface tension near a planar wall, a function of polymer bulk concentration 
$\rho = \rho_{b}$ determined in ref. \cite{JCP116} for SAW polymers; $D_{s}(\rho)$ is the range of the depletion potential
given, according to ref. \cite{JCP117}, by:
\begin{equation}
\label{range.eq}
D_{s}(\rho) = \frac{2 \gamma_{w}(\rho)}{\Pi (\rho)} \frac{R_{AO}^{eff}(q)}{R_{AO}^{eff}(q \rightarrow 0)}
\end{equation}
\noindent
where $\Pi (\rho)$ is the osmotic pressure of the interacting polymers taken from renormalization group calculations
\cite{oono}, namely:
\begin{subequations}
\begin{equation}
\Pi(\rho) = \rho \, Z(\alpha \phi)
\end{equation}
where $\alpha = 2.55$, $\phi = 4 \pi \rho R_{g}^{3}/3$, $R_{g}$ is the polymer radius of gyration at zero density 
($\rho=0$), and
\begin{equation}
Z(x) = 1 + \frac{x}{2} {\rm exp} \left[ 0.309 \times [ (1-\frac{1}{x^{2}}) {\rm ln}(1+x) + \frac{1}{x}] \right]
\end{equation}
\end{subequations}

The second fraction on the r.h.s of (\ref{range.eq}) accounts for finite curvature of the colloid surface, as estimated
approximately from the AO model for ideal polymers \cite{JCP116}:
\begin{equation}
\label{raoeff.eq}
\frac{R_{AO}^{eff}(q)}{R_{g}} = \frac{1}{q}\left(\left(1+\frac{6}{\sqrt\pi}q+3q^{2}\right)^{\frac{1}{3}} - 1 \right) \; , 
\end{equation}
\noindent 
which reduces to $2/\sqrt{\pi}$ in the limit $q \rightarrow 0$.
\subsection{Depletion potential for ideal polymers}
For comparative purposes, we also consider the depletion pair potential induced by ideal polymers.
Meijer and Frenkel \cite{meijer} showed that to a good approximation this is well represented by
the Asakura-Oosawa form \cite{AO}:
\begin{equation}
\label{ao.eq}
\beta v_{id}(r) =
\begin{cases}
- \frac{4 \pi}{3} \rho \sigma^{3} \left\{1- \frac{3}{4} \frac{r}{\sigma}
+ \frac{1}{16} \left(\frac{r}{\sigma}\right)^{3}\right\} & ; \; 2 \, R_{c}  \le  r  \le \;
2 \sigma,  \\
0 &  {\rm otherwise}
\end{cases}
\end{equation}
\noindent 
where $\sigma=R_{c}+R_{AO}^{eff}$ is the radius of a sphere around the
colloids from which the interpenetrable polymers are excluded, but
with an effective $R_{AO}^{eff}$ calculated from the insertion free
energy of a single colloid in a bath of ideal polymer. The radius is
given by eq. (\ref{raoeff.eq}) \cite{JCP116}; for a hard wall it
reduces to $2/\sqrt{\pi} \approx 1.128$, while it monotonically
decreases for increasing size ratio $q$ since the polymer can deform
around the spherical colloid. For the size ratios considered here the
curvature effects are small, on the order of a few $\%$
\cite{aarts}.\\

\begin{figure}[here]
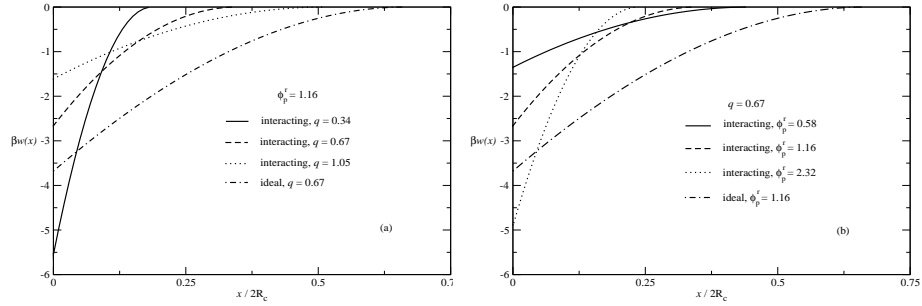

\begin{center}
\includegraphics[width=6cm,height=4cm]{potentielsphi1.16.eps}
\includegraphics[width=6cm,height=4cm]{potentielsq0.67.eps}
\end{center}
\caption{Depletion potential between two colloids as a function of the
surface to surface distance, shown for interacting polymers ($w=v_{s}$
from eq. (\ref{louis.eq})) characterised by $q=0.34, 0.67$ and $1.05$,
for $\phi_{p}^{r} = 1.16$ (a), and by $\phi_{p}^{r} = 0.58, 1.16$
and $2.32$ for $q=0.67$ (b).
The dashed lines indicate the ideal potential
($w=v_{id}$ from eq. (\ref{ao.eq})) for the intermediate cases.  
For a given $\phi_p^{r}$ and $q$, $v_{id}(r)$ is always more
attractive than $v_s(r)$. }
\label{potentiels.fig}
\end{figure}

The two pair potentials (\ref{louis.eq}) and (\ref{ao.eq}) are always
attractive. For interacting polymers the range decreases with
increasing polymer concentration $\rho$, while the latter is constant
for ideal polymers. Furthermore, at any given $\rho$, the depth of
$v_{s}(r)$ is always less than that of $v_{id}(r)$, so that the
depletion attraction induced by interacting polymers is weaker than
for ideal polymers. Representative examples of the depletion
potentials (\ref{louis.eq}) 
are shown in figure~\ref{potentiels.fig}.

\section{Free energy calculations}

Our objective is to draw phase diagrams of \col / \pol mixtures in the ($\eta_{c}$,$\phi_{p}^{r}$) plane, where
$\eta_{c}=4\pi\rho_{c}R_{c}^{3}/3$ is the \col packing fraction, while $\phi_{p}^{r}=4\pi\rho_{p}^{r}R_{g}^{3}/3$ is the 
ratio of the polymer density in the reservoir ($r$) over the overlap density $\rho_{p}^{*}=3/4\pi\rg^{3}$; the latter 
conventionally separates the dilute ($\rho_{p} < \rho_{p}^{*}$) and the semi-dilute ($\rho_{p} > \rho_{p}^{*}$) 
regimes. To this end we need to calculate the free energies of the various phases. Within the semi-grand canonical
ensemble the required free energy is related to the Helmholtz free energy $F_{c}(N_{c},N_{p},V,T)$
by a Legendre transformation:
\begin{equation}
F(N_{c},\phi_{p}^{r},V,T) = F(N_{c},\mu_{p},V,T) = F_{c}(N_{c},N_{p},V,T) - \mu_{p} N_{p}
\end{equation}
\noindent
Since the colloidal particles interact via a hard
sphere repulsion and an effective, depletion-driven attraction, the natural way forward is to calculate the free energies
of the various phases from thermodynamic perturbation theory, using the well-documented hard sphere fluid as a reference 
system \cite{simplliq}. To second order in the high-temperature expansion:
\begin{eqnarray}\label{pert.eq}
F &=& F_{0} + F_{1} + F_{2} \nonumber \\
  &=& F_{0} + \left< W_{N_{c}} \right>_{0} + \frac{\beta}{2} 
\left[ \left< W_{N_{c}}^{2} \right>_{0} - \left< W_{N_{c}} \right>_{0}^{2} \right]
\end{eqnarray}
\noindent 
where $F_{0} = F_{0}(N_{c},V,T)$ is the free energy of the hard sphere fluid, $\left<...\right>_{0}$ denotes an average
over the reference system configurations, and $W_{N_{c}}$ is the perturbation potential energy:
\begin{equation}
W_{N_{c}} = \sum_{i<j}^{N_{c}}w(r_{ij})
\end{equation}
\noindent
with $r_{ij} = |\vec{r}_{i} - \vec{r}_{j}|$, the distance between the centres of colloids $i$ and $j$; 
$w=w(r;\phi_{p}^{r})$ is the \pol concentration-dependent \dep potential (\ref{louis.eq}) for \inter polymers ($w=v_{s}$)
or (\ref{ao.eq}) for non-\inter polymers ($w=v_{id}$). We stress that within the semi-grand canonical description the 
depletion potential must be calculated for the polymer density in the reservoir, which is unequivocally determined by
fixing the chemical potential $\mu_{p}$.

$F_{0}$ in eq. (\ref{pert.eq}) is calculated from the Carnahan and
Starling equation of state for the \hs fluid \cite{CS} while for the
\hs solid we adopt Hall's equation of state \cite{hall}. $F_{1}$ is
easily expressed in terms of the \hs pair distribution function
$g_{0}(r)$ for which we adopted the Verlet-Weis parametrisation in the
fluid \cite{VW}, and the form proposed by Kincaid and Weis for the FCC
solid phase \cite{KW}. The calculation of $F_{2}$ involves three and
four-body contributions of the reference system. We have adopted the
approximate expression due to Barker and Henderson \cite{barkerhend},
which only involves the pair distribution function and the
compressibility of the reference system. Gathering results:
\begin{equation}
\label{perturb.eq}
\frac{\beta F}{N_{c}} = \frac{\beta F_{0}}{N_{c}} + \frac{1}{2} \beta \rho_{c} \int d^{3} r \, g_{0}(r) \, w(r)
- \frac{1}{4} \left( \frac{\partial \rho_{c}}{\partial p} \right)_{0} \beta \rho_{c} 
\int d^{3} r \, g_{0}(r) \, w^{2}(r) \, .
\end{equation}
\noindent 
Note that in the solid phase, $g_{0}(r)$ is the orientationally-averaged pair distribution function.

In order to assess the accuracy of $F_{2}$ and the convergence of the perturbation series (\ref{perturb.eq}), we have 
carried out MC simulations to compute explicitly the fluctuation term in eq. (\ref{pert.eq}), which is approximated by
the last term in eq. (\ref{perturb.eq}), as well as the total excess free energy. The latter is most conveniently 
calculated by the standard $\lambda$-integration procedure \cite{simplliq}, whereby the depletion-induced perturbation
$W_{N_{c}}$ is gradually switched on, resulting in:
\begin{equation}\label{lambdaint.eq}
F(\lambda=1) = F(\lambda=0) + \int_{0}^{1}\left< W_{N_{c}} \right>_{\lambda} d\lambda
\end{equation}
\noindent
where $F(\lambda=1)$ is the required free energy of the fully interacting \col / \pol mixture, $F(\lambda=0) \equiv 
F_{0}$ is the free energy of the \hs reference system, and $\left< W_{N_{c}} \right>_{\lambda}$ is the statistical
average of the perturbation weighted by the Boltzmann factor appropriate for a system of particles \inter via the \hs
repulsion and the partially switched on \dep potential $\lambda w(r)$. The calculation of the free energy $F$ hence
involves several MC simulations to determine $\left< W_{N_{c}} \right>_{\lambda}$ for a series of discrete values of
$\lambda$\cite{PR184}, typically $\lambda = n \times 0.05$ ($1 \leq n \leq 20$).

\begin{figure}[here]
\begin{center}

\includegraphics[width=15cm,height=7.5cm]{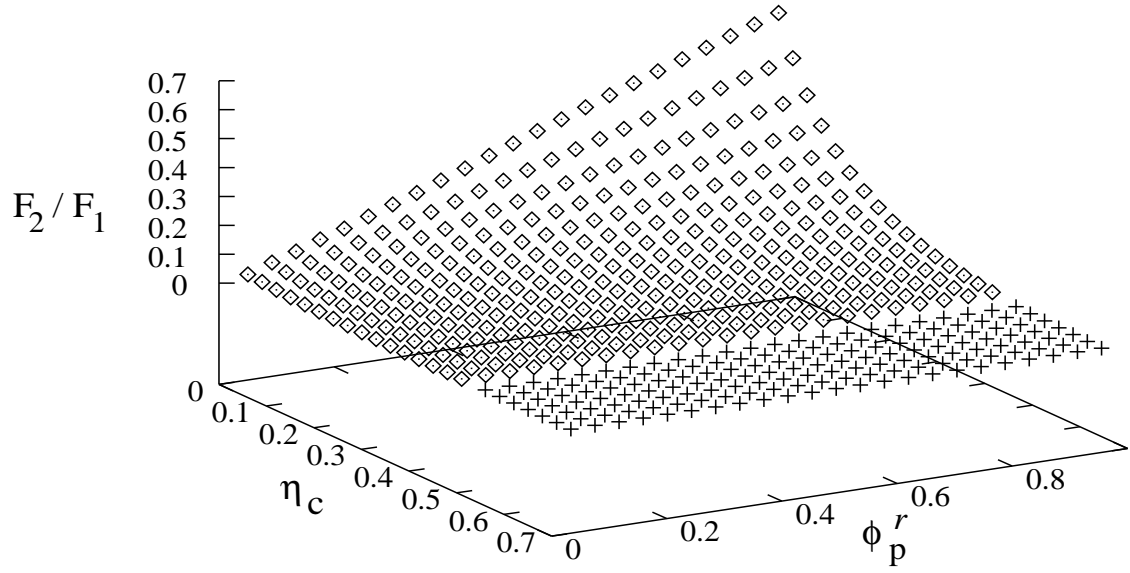}
\end{center}
\caption{Ratio of the first two perturbative terms $F_{2}/F_{1}$, as a
function of the colloid packing fraction $\eta_{c}$ and the polymer
density in the reservoir $\phi_{p}^{r}$. The diamonds are for the
fluid phase and the crosses ($\eta_c \geq 0.5$) for the solid
phase. The size ratio is $q = 0.67$.}
\label{f1f2color.fig}
\end{figure}

The convergence of the perturbation series (\ref{perturb.eq}) is illustrated in figure \ref{f1f2color.fig} 
for the \dep potential (\ref{louis.eq}) and a size ratio $q=0.67$. As expected, the convergence of the
series is faster when the size ratio $q$ is larger, the polymer concentration is lower and the \col packing 
fraction higher.\\

The accuracy of each term in the series, and of the truncated sum,
were tested by MC simulations of periodic samples of $N_{c}=108$
colloidal particles. Representative results are presented in tables I
and II. Table I shows that values of $F_{1}$ from
eq. (\ref{perturb.eq}) are very close to the simulation data, both for
the fluid and the solid. The Barker-Henderson approximation
underestimates the absolute value of $F_{2}$, by less than a factor of
two in the fluid phase, and by a much larger factor in the FCC crystal
phase, where it is totally inadequate. However, as shown in Table II,
the sum of the first three terms of the perturbation series
(\ref{perturb.eq}) yields a total free energy which is surprisingly
close to the ``exact'' MC results from the
$\lambda$-integration. Similar comparisons show that the predictions
of perturbation theory are very reliable both for larger and for
somewhat smaller values of $q$ (say for $q \gtrsim 0.3$), but that the
predictions rapidly deteriorate for small $q$, corresponding to narrow
potential wells, as expected. This failure will be illustrated in the
case $q=0.1$ at the end of the following section.

\section{Phase diagrams}

\begin{figure}[here]
\begin{center}
\includegraphics[width=14cm,height=10cm]{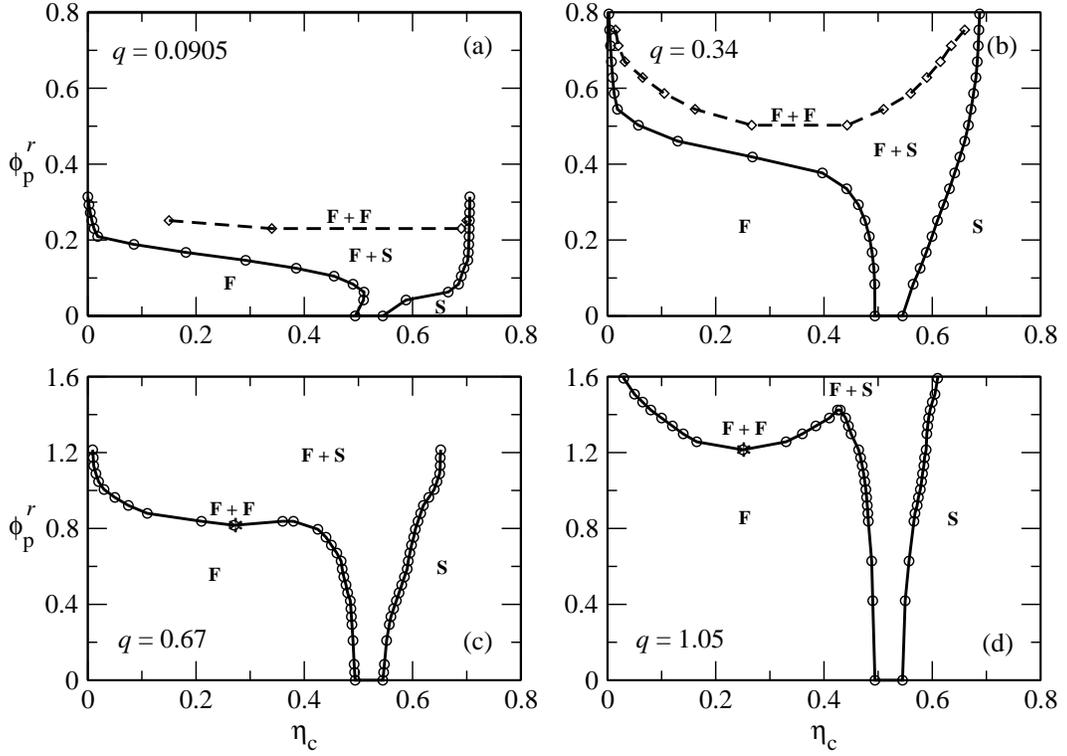}
\end{center}
\caption{Phase diagrams of colloid / interacting polymer mixtures as
obtained from perturbation theory for the effective one component
system with size ratios (a) $q=R_{g}/R_{c}=0.0905$, (b) $q=0.34$, (c)
$q=0.67$, and (d) $q=1.05$, in the plane of the colloid packing
fraction $\eta_{c}$ and the polymer concentration $\phi_{p}^{r}$ in
the reservoir. F and S denote the stable fluid and solid (FCC)
phases. F + S and F + F denote the stable fluid-solid and (meta)stable
fluid-fluid coexistence regions. The solid lines denote the phase
boundaries for the coexistence of stable phases, while the dashed
lines denote the metastable fluid-fluid binodal.  The critical points
are indicated by asterisks.}
\label{pert-int.fig}
\end{figure}

Once the free energies of the fluid and FCC solid phases have been
calculated from thermodynamic perturbation theory, as explained in the
previous section, the phase diagrams can be calculated using the
standard double-tangent construction. Since the initial two-component
system is athermal, the \dep potentials (\ref{louis.eq}) and
(\ref{ao.eq}) are purely entropic, so that the temperature scales out
in the Boltzmann factor, and the resulting phase diagrams are
independent of temperature. The phase diagrams for mixtures of
colloids and \inter polymers in the ($\eta_{c}$,$\phi_{p}^{r}$) plane
are shown in figure \ref{pert-int.fig} for four values of the size
ratio $q$, in the range $0.1 \lesssim q \lesssim 1$.  The phase
diagrams look superficially similar to earlier results obtained for
mixtures of colloids and ideal polymers \cite{lekk,dijkstra} or star
polymers \cite{dzubiella}. In particular, for the smaller size ratios,
the fluid-fluid phase separation is metastable, and preempted by phase
coexistence between a high density solid and a single low density
fluid phase. Such a behaviour is a typical signature of ``narrow''
potential wells like those pictured in figure \ref{potentiels.fig}
\cite{gast,lekk}.  For larger size ratios a stable fluid-fluid phase
separation appears with a critical point and a triple point, and the
resulting phase diagrams are not unlike those of simple atomic systems
in the density-temperature plane (with $1/\phi_{p}^{r}$ playing the
role of $T$).

\begin{figure}[here]
\begin{center}
\includegraphics[width=14cm,height=10cm]{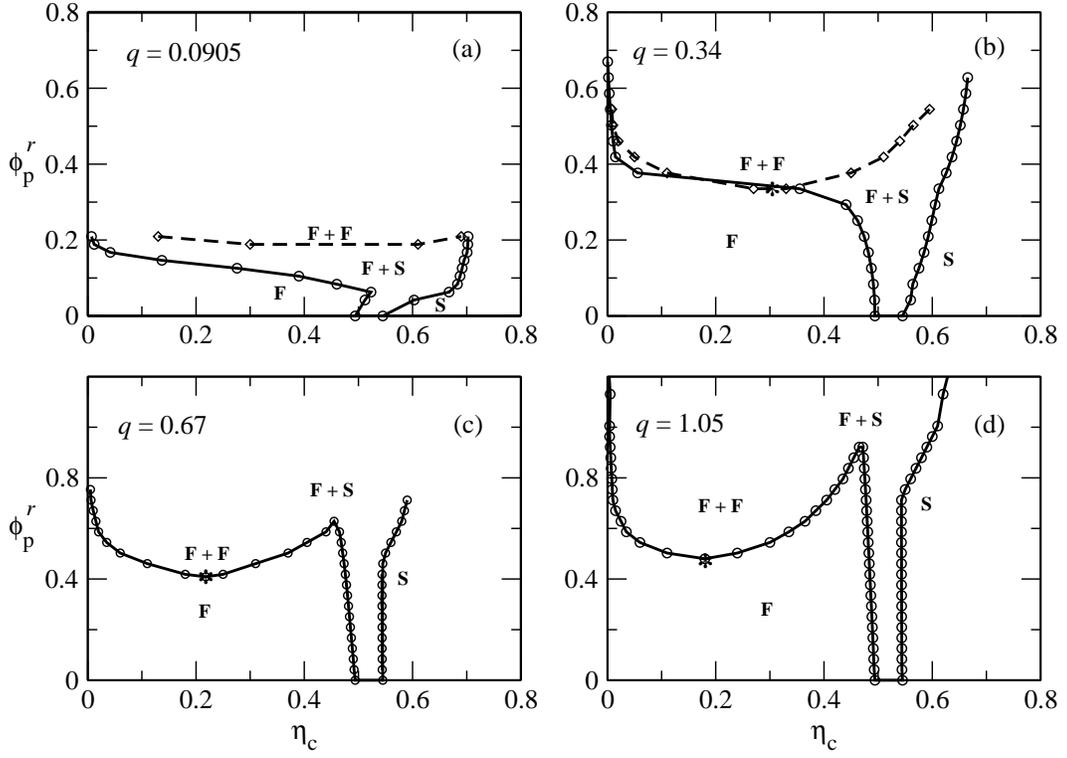}
\end{center}
\caption{Phase diagrams of colloid / ideal polymer mixtures as obtained from perturbation theory
for the effective one-component system using $v_{id}(r)$ (eq. (\ref{ao.eq})) for the same $q$'s
as in figure \ref{pert-int.fig},
as functions of the colloid packing fraction $\eta_{c}$ and the polymer
concentration $\phi_{p}^{r}$ in the reservoir.
Note that the critical points (asterisks) are always at lower $\phi_{p}^{r}$ 
than the corresponding crtitical points for interacting polymers.}
\label{pert-AO.fig}
\end{figure}

The corresponding phase diagrams for mixtures of colloids and ideal
polymers, calculated using the ideal \dep potential (\ref{ao.eq}) are
shown figure \ref{pert-AO.fig} for comparison. While they look
qualitatively similar to those for \inter polymers in figure
\ref{pert-int.fig}, there are a number of striking quantitative
differences. Because the \dep attraction for ideal polymers
(eq. \ref{ao.eq}) is stronger than that for \inter polymers
(eq. (\ref{louis.eq})) for the same \pol concentration, the
fluid-fluid phase separation becomes stable at a larger $\phi_{p}^{r}$
for the \inter than for the non-\inter polymers.  While the phase
diagrams for $q \simeq 0.1$ are fairly close, the differences grow
with increasing $q$. For $q \simeq 1$ the critical point in figure
\ref{pert-AO.fig} (ideal case) is at
($\eta_{c}=0.18$,$\phi_{p}^{r}=0.48$) compared to (0.25,1.21) in the
\inter case, while the triple points are at (0.47,0.92) and
(0.43,1.42) respectively, indicating dramatic changes when going from
ideal to \inter polymers. Also note that while the critical polymer
concentration is practically independent of $q$ (for $q \gtrsim 0.35$)
in the ideal case, it shifts to higher values as $q$ increases in the
\inter case. On the other hand, the critical \col packing fraction
decreases as $q$ increases in the non-\inter case, while it is
practically constant for \inter polymers.

\begin{figure}
\begin{center}
\includegraphics[width=14cm,height=10cm]{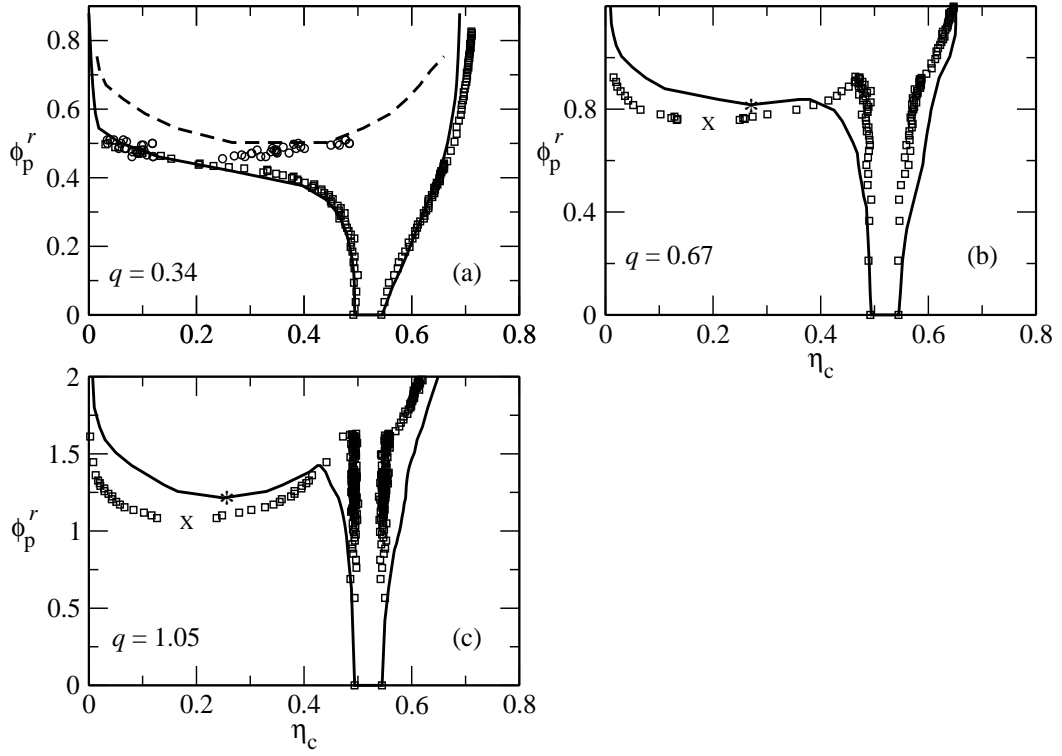}
\end{center}
\caption{Phase diagrams of colloid / interacting polymer mixtures, as
obtained from perturbation theory for the effective one-component
system (lines), compared to simulations of the two-component system
(symbols) \protect\cite{PRL89}.  The size ratios are (a) $q = 0.34$,
(b) $0.67$ and (c) $1.05$.  The critical points are indicated by
crosses (two-component) and asterisks (one-component).  The agreement
is surprisingly good.}
\label{pert-sim22.fig}
\end{figure}

All these trends are similar to those reported recently in simulations
of the two-component description of mixtures of colloids and SAW
polymers \cite{PRL89}. A detailed comparison between the present
perturbation theory results for the effective one-component system,
and the phase diagrams determined for the two-component representation
is made in figure \ref{pert-sim22.fig}. The agreement between the
simulation data for the two-component representation and the
predictions of perturbation theory for the effective one-component
representation is seen to be reasonable, but not perfect, and to
deteriorate as $q$ increases. The obvious reason is that perturbation
theory only includes the pair-wise additive part of the \dep
interactions, while the two-component representation also accounts for
effective many-body \dep interactions between colloidal particles. The
fact that the phase diagrams obtained from the two-component
representation are shifted to lower polymer concentrations relative to
the prediction for the effective one-component system indicates that
the more-than-two-body \dep interactions are overall attractive in
nature\footnote{Of course part of the difference is also due to the
perturbation theory, which, in general, slightly underestimates the
value of $\phi_p^r$ along phase boundaries (see e.g.\ the work of
Dijkstra {\em et al.}\protect\cite{dijkstra}). This suggests that the
many-body interactions are slightly more attractive than would be
inferred from figure~\protect\ref{pert-sim22.fig}. }. The opposite trend
was found in the case of ideal polymers \cite{dijkstra}, and is
consistent with the more pronounced trends in the limit of large $q$
\cite{long-pol}.

\begin{figure}[here]
\begin{center}
\includegraphics[width=14cm,height=10cm]{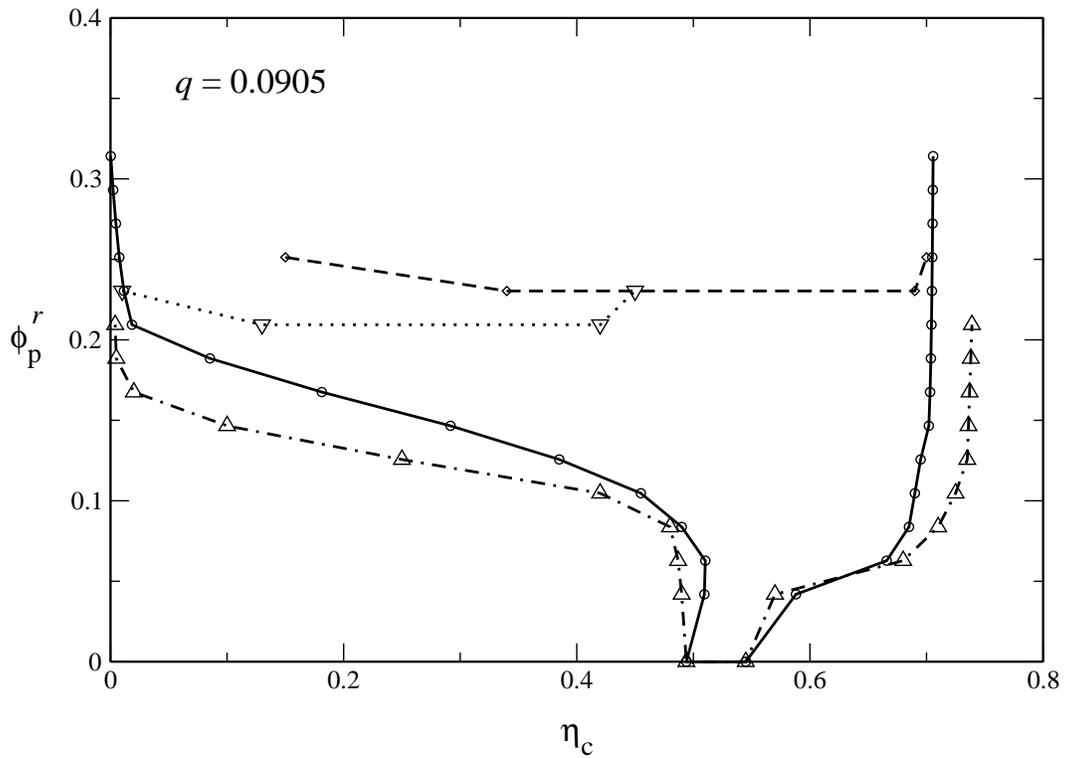}
\end{center}
\caption{Phase diagram for colloid / interacting polymer mixture, as a
function of the colloid packing fraction $\eta_{c}$ and the polymer
concentration $\phi_{p}^{r}$ in the reservoir, for a small size ratio
($q=0.0905$).  The solid (resp. dashed-dotted) line denotes the
fluid-solid binodal obtained from perturbation theory (resp. from MC
simulations of the effective one-component system), while the dashed
(resp. dotted) line denotes the metastable fluid-fluid binodal
obtained by the same methods.}
\label{pert-sim1.fig}
\end{figure}

We pointed out earlier that the convergence of thermodynamic
perturbation theory is expected to deteriorate when the range of the
attractive potential well decreases, {\it i.e.} when $q$ decreases. To
check the reliability of second order perturbation theory at $q \simeq
0.1$, we have systematically computed the ``exact'' free energy by MC
simulations, using the $\lambda$-integration
(eq. (\ref{lambdaint.eq})). The phase diagrams determined with the
approximate and ``exact'' free energies are compared in figure
\ref{pert-sim1.fig}.  The agreement remains acceptable for the
fluid-solid transition, even for $q \simeq 0.1$, but the (metastable)
critical point of the fluid-fluid transition is at too high a colloid
packing fraction\cite{dijkstra}.  

Two-component simulations\cite{PRL89} would be very expensive for small $q$,
because the number of polymers needed scales as $q^{-3}$. However, for
sufficiently small $q$ we don't expect many-body interactions to be
important, and so our one-component simulation should accurately
represent the true colloid / polymer system.

\begin{figure}[here]
\begin{center}
\includegraphics[width=14cm,height=10cm]{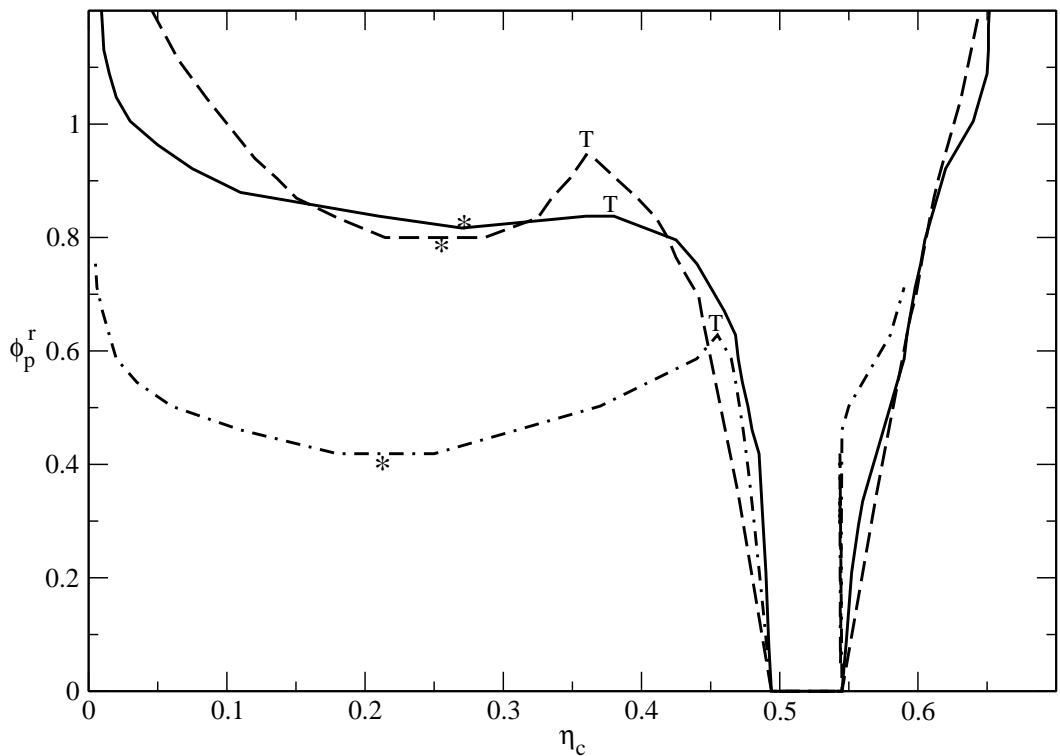}
\end{center}
\caption{Phase diagrams of colloids/polymer mixtures within three different one-component models: 
ideal polymers (dashed-dotted lines, $R_{g} = 0.67$), interacting polymers (solid lines, $R_{g} = 0.67$) 
and star polymers with functionality $f=2$ (dashed lines, $R_{g} = 0.6$). The critical points are indicated by
asterisks, while the symbol $T$ denotes the triple points.}
\label{compstarf2.fig}
\end{figure}

A final instructive comparison is between the present results for the
phase behaviour of colloid/interacting polymer mixtures and the
results for colloid/star polymer mixtures of functionality
$f=2$\cite{dzubiella}, which reduce in fact to interacting linear
polymers considered in the present work.  The phase diagrams
calculated from both depletion potentials within the same
approximation (\ref{perturb.eq}) for the free energies are compared in
figure \ref{compstarf2.fig} for similar size ratios $q$.  Although, as
explained earlier in this paper, the depletion pair potential
calculated for the $f=2$ star polymers \cite{dzubiella} is not quite
the same as the more accurate one we use\cite{JCP117}, the
phase-diagrams show the similar trends when compared to ideal
polymers.

\section{Conclusion}
We have shown that traditional thermodynamic perturbation theory,
requiring only the well-documented equations of state and pair
distribution functions of the fluid and solid phases of the reference
hard sphere system, leads to reasonably accurate phase diagrams of
mixtures of colloidal particles and {\it interacting} polymer coils,
provided the appropriate concentration-dependent depletion potential
between two colloidal spheres is used.  As expected, the agreement
between the predictions of the effective one-component description,
and the more elaborate two-component description observed for low size
ratios $q$ deteriorates as $q$ increases, due to the enhanced
importance of many-body interaction which are neglected in the
one-component picture. Nevertheless, the disagreement remains tolerable
even at $q\approx 1$, and in view of the excellent agreement between
the predictions of the two-component description \cite{PRL89} and
recent experimental data \cite{ilett,ramak}, we conclude that the
effective one-component picture, in conjunction with standard
thermodynamic perturbation theory, provides a reliable prediction of
the phase diagrams of colloids/polymer mixtures in good solvent.

A direct comparison between the phase diagrams for interacting and
ideal polymers calculated at the same level of approximation shows
considerable quantitative, and even qualitative differences between
the two depletants.  The main effect of polymer-polymer interactions
is to enhance the miscibility of the colloid / polymer mixtures.
Similar conclusions were reached by a number of different recent
investigations, based on two-component approaches,
including integral equations\cite{fuchs}, ``polymers
as soft colloids''\cite{PRL89}, extensions of free-volume
theory\cite{aarts}, density functional theory\cite{schmidt}, and
star-polymer potentials\cite{dzubiella}.  Here we show that the
 differences between the two types of depletants can be rationalised
within a one-component effective potential picture, mainly because for
a given $R_g$ and density $\rho_p$, the depletion potentials for
interaction polymers are less attractive than those for interacting
polymers.

The results of the present work apply to polymers in good solvent, for
which the SAW model constitutes an excellent representation. We
plan to examine the situation where solvent quality is such that
attractive forces between monomers can no longer be neglected
\cite{krakovi}. 
Upon lowering the temperature from very high
(corresponding to the SAW limit) to the $\theta$ temperature, we
should be able to investigate the gradual change in the phase diagrams
from the fully interacting case to one similar to the ideal polymer limit, which have
both been considered in the present paper.

\acknowledgments 
The financial support of the EPSRC (grant number RG/R70682/01) is gratefully acknowledged. 
A.A. Louis acknowledges financial support from the Royal Society.
B. Rotenberg acknowledges financial support from the Ecole Normale Sup\'erieure de Paris.

\newpage

\begin{table}\label{table1.tab}
\begin{center}
\begin{tabular}{|c|c|c|c|c|c|c|} \cline{4-7}
\multicolumn{3}{c}{} &\multicolumn{2}{|c|}{Perturbations} &\multicolumn{2}{|c|}{Simulations} \\ \hline
$\phi_{p}^{r}$ & $\eta_{c}$ & State & $\widetilde{F}_{1}^{pert}$ & 
$\widetilde{F}_{2}^{pert}$ & $\widetilde{F}_{1}^{MC}$ & $\widetilde{F}_{2}^{MC}$\\ \hline  \hline 
0.17           &    0.22    & Fluid &     -0.194     &     -0.004     &    -0.193    &   -0.006    \\ \cline{2-7}
               &    0.64    & Solid &     -2.692     &     -0.001     &    -2.701    &   -0.032    \\ \hline \hline
0.29           &    0.22    & Fluid &     -0.318     &     -0.012     &    -0.318    &   -0.021    \\ \cline{2-7}  
               &    0.64    & Solid &     -4.663     &     -0.003     &    -4.679    &   -0.097    \\ \hline \hline
0.42           &    0.22    & Fluid &     -0.434     &     -0.025     &    -0.433    &   -0.043    \\ \cline{2-7}
               &    0.64    & Solid &     -6.900     &     -0.006     &    -6.941    &   -0.210    \\ \hline \hline      
0.54           &    0.22    & Fluid &     -0.535     &     -0.041     &    -0.534    &   -0.077    \\ \cline{2-7}
               &    0.64    & Solid &     -8.604     &     -0.011     &    -8.635    &   -0.329    \\ \hline
\end{tabular}
\caption{
First two terms of the perturbation series, as obtained using equation (\ref{perturb.eq}) ($F^{pert}$),
and by MC simulations ($F^{MC}$), for several colloid and polymer concentrations, and for $q=0.67$.
The free energy densities are given in reduced units $\widetilde{F} = \beta F (2 R_{c})^{3} / V$. Whereas both 
$F_{1}^{pert}$ and $F_{2}^{pert}$ are accurate in the fluid phase, the latter underestimates the absolute value of 
$F_{2}^{MC}$ in the solid phase.}
\end{center}
\end{table}

\begin{table}\label{table2.tab}
\begin{center}
\begin{tabular}{|c|c|c|c|c|c|} \cline{4-6}

\multicolumn{3}{c}{} &\multicolumn{2}{|c|}{$\widetilde{F}_{0}+\widetilde{F}_{1}+\widetilde{F}_{2}$} &   $\widetilde{F}$\\ \hline
$\phi_{p}^{r}$ & $\eta_{c}$ & State & Perturbations  &  Simulations   & $\lambda$-integration \\ \hline \hline
0.17           &    0.22    & Fluid &     -0.055     &     -0.056     &    -0.056      \\ \cline{2-6}
               &    0.64    & Solid &      7.551     &      7.511     &     7.541    \\ \hline \hline
0.29           &    0.22    & Fluid &     -0.194     &     -0.196     &    -0.196      \\ \cline{2-6}  
               &    0.64    & Solid &      5.578     &      5.468     &     5.562      \\ \hline \hline
0.42           &    0.22    & Fluid &     -0.316     &     -0.333     &    -0.339      \\ \cline{2-6}
               &    0.64    & Solid &      3.582     &      3.368     &     3.563      \\ \hline \hline      
0.54           &    0.22    & Fluid &     -0.434     &     -0.468     &    -0.483      \\ \cline{2-6}
               &    0.64    & Solid &      1.629     &      1.279     &     1.605      \\ \hline
\end{tabular}
\end{center}
\caption{
Truncated free energy densities $F_{0}+F_{1}+F_{2}$, in reduced units (see table I),
as obtained using equation (\ref{perturb.eq}) and MC simulation,
and total free energy of the system, as obtained from $\lambda$-integration (equation (\ref{lambdaint.eq})). Whereas
the results of MC simulations are accurate only in the fluid phase, thoses obtained using equation (\ref{perturb.eq})
are in good agreement with $\lambda$-integration in both phases.}
\end{table}

\clearpage
\newpage

\clearpage
\newpage

\end{document}